\documentclass[aps,prx,twocolumn,amsmath,amssymb,superscriptaddress,floatfix,longbibliography]{revtex4-1}

\usepackage{graphicx}
\usepackage{multirow}
\usepackage{color}
\usepackage{bm}
\usepackage{ulem}

\renewcommand{\Vec}[1]{\mbox{\boldmath$#1$}}
\def\infinity{\infty}
\def\t#1{\textrm{#1}}
\def\ket#1{|#1\rangle }
\def\bra#1{\langle #1 |}
\def\bracket#1{\langle #1 \rangle}
\def\n{\nonumber \\ }

\begin{document}

\title{
Topological magneto-electric effects in
thin films of topological insulators
}

\author{Takahiro Morimoto}
\affiliation{RIKEN Center for Emergent Matter Science 
(CEMS), Wako, Saitama, 351-0198, Japan}
\author{Akira Furusaki}
\affiliation{Condensed Matter Theory Laboratory, 
RIKEN, Wako, Saitama, 351-0198, Japan}
\affiliation{RIKEN Center for Emergent Matter Science 
(CEMS), Wako, Saitama, 351-0198, Japan}
\author{Naoto Nagaosa}
\affiliation{RIKEN Center for Emergent Matter Science 
(CEMS), Wako, Saitama, 351-0198, Japan}
\affiliation{Department of Applied Physics, The University of 
Tokyo, Tokyo, 113-8656, Japan}

\date{\today}

\begin{abstract}
We propose that the topological magneto-electric (ME) effect,
a hallmark of topological insulators (TIs),
can be realized in thin films of 
TIs in the $\nu=0$ quantum Hall state under magnetic field
or by doping two magnetic ions with opposite signs of exchange coupling.
These setups have the advantage compared to previously proposed setups
that a uniform configuration of magnetic field or magnetization
is sufficient for the realization of the topological ME effect.
To verify our proposal, we numerically calculate ME response of
TI thin films in the cylinder geometry and that of
effective 2D models of surface Dirac fermions.
The ME response is shown to converge to the quantized value 
corresponding to the axion angle $\theta=\pm \pi$
in the limit of the large top and bottom
surface area of TI films,
where
non-topological contributions from the bulk and the side surface
are negligible.
\end{abstract}

\maketitle

\section{Introduction}
Topological aspects of the condensed matter systems 
have been a subject of intensive studies
since the discovery of  
the quantum Hall effect (QHE)~\cite{QHE1,QHE2}.
The topological nature of the QHE is encoded in
the Hall conductance, which is exactly quantized in units of 
$e^2/h$ 
to the Chern number
characterizing the global topology of wave functions constituting a
manifold in the Hilbert space. 
The Hall current supporting the
quantized Hall conductance flows through the edge channels
along the boundary of the sample (bulk-edge correspondence). 

The extension of the QHE
to time-reversal symmetric systems is realized in 
topological insulators (TIs)~\cite{TI1,TI2}.
Here, 
the $\mathbb{Z}_2$ topological index
characterizes the
global topology of wave functions of Kramers pairs.
Helical edge channels and surface Dirac fermions
exist on the boundary of two-dimensional (2D) and
three-dimensional (3D) TIs, respectively,
as a consequence of nontrivial topology. 
The spin Hall conductance of 2D TIs
is an analog of the Hall conductance in the QHE. 
However, the spin current is not
conserved in the presence of spin-orbit interactions
which are indispensable for TIs; the spin Hall conductance is neither
a well-defined nor quantized quantity.
The charge conductance along the helical edge channels
in 2D TIs was shown to be quantized at $2e^2/h$~\cite{molenkamp},
but the quantization is not so accurate as in the QHE.
Furthermore, one cannot relate the charge conductance to
the $\mathbb{Z}_2$ topological index in 3D TIs.
The quantized quantity characteristic of TIs, 
corresponding to the Hall conductance in the QHE, has been looked for.

The topological magneto-electric (TME) effect%
~\cite{TME,Essin09,Essin10,Qi09,Tse,Maciejko,Malashevich,Mong,Nomura,Hwang12}
has been suggested as a phenomenon
that is directly related to
the $\mathbb{Z}_2$ topological index for 3D TIs.
The TME effect is a quantized response of polarization to applied magnetic
fields or a quantized response of magnetization to applied electric
fields~\cite{TME,Essin09}. 
The quantized polarization current was proposed to be accessible
by transport measurements \cite{TME,Essin09,Essin10}.
Alternatively, the TME effect was proposed to be observed by
optical measurements of Faraday or Kerr rotation showing
a quantized response~\cite{Tse,Maciejko}.
However, no experimental observation of the TME effect has
been reported so far.

The TME effect of a 3D TI is theoretically described
by the axion action (with $c=1$),
\begin{equation}
S
=\frac{\theta e^2}{32 \pi^2 \hbar} \int dt d^3x
\epsilon^{\mu\nu\lambda\delta}F_{\mu\nu}F_{\lambda\delta}
=\frac{\theta e^2}{4 \pi^2 \hbar} \int dt d^3x
\Vec{E} \cdot \Vec{B},
\label{eq:theta}
\end{equation}
where the angle $\theta$ is equivalent to the $\mathbb{Z}_2$ index:
$\theta=\pi~(\t{mod } 2\pi)$ for TIs and
$\theta=0~(\t{mod } 2\pi)$ for trivial insulators \cite{TME}.
Transforming Eq.~(\ref{eq:theta}) to the integral over the
surface of a 3D TI
yields the Chern-Simons action,
\begin{equation}
S_\t{CS}
=\frac{\theta e^2}{8 \pi^2 \hbar} \int dt d^2x
\epsilon^{\mu\nu\lambda}A_{\mu}\partial_\nu A_{\lambda},
\label{eq:CS}
\end{equation}
which describes the QHE on the TI surface
with the Hall conductance 
$\sigma_{xy} = (\theta/2\pi) (e^2/h)$.
For TIs with $\theta=\pi$, this amounts to 
$\sigma_{xy} = e^2/2h$.
Microscopically, the half-integer QHE
($\sigma_{xy} = \pm e^2/2h$) on the TI surface
occurs once surface Dirac fermions acquire a mass
and have the Hamiltonian
\begin{equation}
H=v_F(-p_x \sigma_y + p_y \sigma_x) + M \sigma_z,
\label{eq:Weyl}
\end{equation}
where $v_F$ is the Fermi velocity.
The mass gap $M$ of the surface Dirac fermions can be introduced,
for example, through the exchange coupling $J$ of Dirac fermions to
the magnetization $m$ (pointing perpendicular to the surface)
of doped ferromagnetic ions; $M=Jm$.
In order to realize the TME effect described by
Eqs.~(\ref{eq:theta}) and (\ref{eq:CS}),
two conditions must be satisfied.
(i) The magnetization points outward or
inward over the whole surface of a 3D TI such that surface Dirac fermions
are fully gapped on any surface. 
(ii) The Fermi energy must be tuned to be
inside the gap induced by the magnetization $m$. 
This magnetic configuration is very difficult to realize 
experimentally since the external magnetic field favors the 
same magnetization direction for both top and bottom surfaces.
In addition, 
the fine tuning of the Fermi energy is still very difficult experimentally although some groups have succeeded~\cite{Chang,Checkelsky2}.
We note in this regard that localization effect can relax the second condition.
That is, when the surface Dirac fermions are localized by 
magnetic impurities, the Fermi energy can be set
in the wider energy window of localized states~\cite{Nomura}. 
Furthermore, the magneto-electric cooling effect can help the magnetization
$m$ align perpendicular to any surface because of the
energy gain proportional to the volume which eventually overcome the
surface energy~\cite{Nomura}.
However, there is no experimental study
to pursue this possibility up to now.

In this paper we propose alternative
routes to realizing
the TME effect in thin films of 3D TIs by either applying external
magnetic field
or doping two kinds of magnetic impurities.
The first approach utilizes the QHE of TI thin films~\cite{Brune,Xu,Yoshimi}.
In particular, we are interested in the novel $\nu=0$ QHE
which was recently observed
in the presence of a finite potential difference between the top and
bottom surfaces \cite{Yoshimi}.
In this case, the edge channels are gapped on the side surfaces
due to the finite thickness effect, which can produce interesting 
spintronics functions~\cite{Morimoto15}.
We point out that this $\nu=0$ quantum Hall system provides
an ideal laboratory to realize the TME effect,
as discussed in detail below.
We also propose that the TME effect can be achieved 
by doping Cr and Mn to upper and lower 
halves of TI thin films, respectively.

\begin{figure}[tb]
\begin{center}
\includegraphics[width=\linewidth]{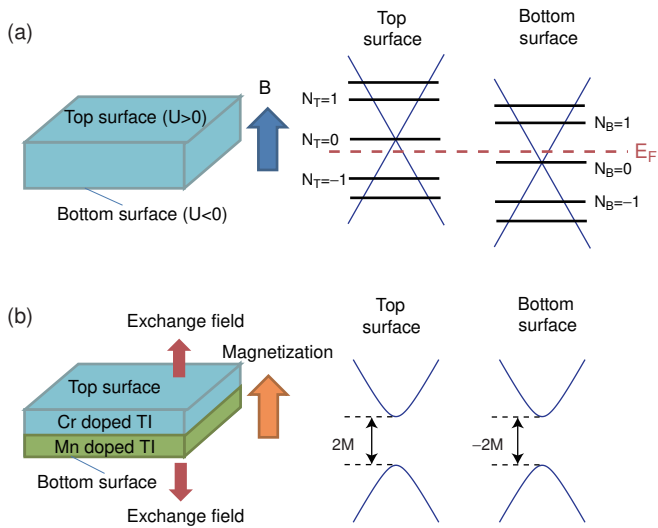}
\end{center}
\caption{
Setups for the TME effect in (a) TI thin films in the $\nu=0$ QHE, and
(b) magnetically doped TI thin films. Right panels are schematics of energy bands of surface states.
}
\label{Fig: TI surface}
\end{figure}

\section{Topological magneto-electric effects in
TI thin films: Basic idea}
The TME effect described by the actions in Eqs.~(\ref{eq:theta})
and (\ref{eq:CS}) predicts that charge polarization $\Vec P$ is
induced by applying a magnetic field $\Vec B$,
\begin{align}
\Vec{P}=\frac{\theta}{2\pi}\frac{e^2}{h} \Vec{B},
\label{P = theta B}
\end{align}
where $\theta=\pi$ (mod $2\pi$).
In this section we explain how
this quantized response of charge polarization is 
obtained for the proposed two setups of TI thin films
by discussing charge response of quantum Hall states
of Dirac fermions at the top and bottom surfaces.

Let us consider a thin film of a TI in the magnetic field
$\bm{B}=(0,0,B)$
which is applied perpendicular to the film; see Fig.~\ref{Fig: TI surface}(a).
In this case, Dirac fermions at the top and bottom surfaces
form Landau levels (LLs).
For simplicity we assume that the Fermi energy is at
the Dirac point, $E_F=0$.
In this case, important low-energy excitations are states in the LLs
at the Fermi energy on the top and bottom surfaces,
while the excitations are gapped in the bulk
and at the side surface (the latter due to
the finite-thickness effect).
In addition, we introduce a potential energy difference
between the top surface ($U_0>0$) and bottom surface ($-U_0$).
For small magnitude of the potential asymmetry $U_0$,
the LLs at the top and bottom surfaces are occupied up to $N_T=-1$
and $N_B=0$ LLs, respectively, as indicated in Fig.~\ref{Fig: TI surface}(a),
where $N_T$ and $N_B$ are LL indices on the top and bottom surfaces.
This situation realizes the $\nu=0$ QHE
which was experimentally observed in Ref.~\cite{Yoshimi}.
More generally,
the $\nu=0$ QHE is realized in TI thin films
when LLs at the top and bottom surfaces are filled up to
$N_T=-N-1$ and $N_B=N$ LLs with any integer $N$,
where $N$ can be controlled by changing 
the potential asymmetry $U_0$.
Then the charge densities at the top and bottom surfaces ($n_T$ and $n_B$) 
become
\begin{align}
n_{T}&=\left(N+\frac{1}{2} \right)\! \frac{e^2}{h} |B| ,
&
n_{B}&=-\left(N+\frac{1}{2} \right)\! \frac{e^2}{h} |B| .
\end{align}
If the top and bottom surfaces are located at $z=d$ and $z=-d$,
respectively,
the charge polarization along the $z$ direction is given by
\begin{align}
P=\frac{1}{2d}[d n_T +(-d)n_B ]
=\left( N+\frac{1}{2}\right)\! \frac{e^2}{h} |B|.
\label{eq: P in nu=0 QHE}
\end{align}
The polarization in the $\nu=0$ QHE
coincides with the value expected 
from the TME effect, Eq.~(\ref{P = theta B})
with $\theta=(2N+1)\pi$,
for $B>0$.
For $B<0$, the axion angle is given by $\theta=-(2N+1)\pi$.
This jump of $\theta$ at $B=0$ takes place 
because the surface bands become gapless at $B=0$. 
In both cases, the axion angle takes a nontrivial value $\theta=\pi$
modulo $2\pi$.
We emphasize again that, in the setup with the $\nu=0$ QHE,
an arbitrary odd integer $\theta/\pi=2N+1$ can be realized 
by tuning the potential asymmetry $U_0$
with gating the top and bottom surfaces independently
such that the highest occupied LLs are set to $N_T=-N-1$ and $N_B=N$.

Next we consider the second setup in which
the upper and lower halves of a TI film are doped with two types of
magnetic ions, e.g., Cr and Mn.
Here a crucial assumption is that
the two kinds of magnetic ions have exchange couplings $J$ of opposite signs
with Dirac fermions in the TI film.
This is indeed the case with Cr and Mn,
as can be seen from the fact that
Cr-doped and Mn-doped $\t{Bi}_2 \t{Te}_3$-based TIs
show quantized anomalous Hall conductivity of opposite signs
when the Fermi energy is tuned within the excitation gap of
surface Dirac fermions
\cite{Checkelsky1,Chang,Checkelsky2}.
Then a uniform magnetization depicted in Fig.~\ref{Fig: TI surface}(b)
produces an exchange field pointing outward or inward
at the top and bottom surfaces, and the desired constant
Dirac mass $Jm\sigma_z$ in Eq.~(\ref{eq:Weyl}) is
achieved except for the side surface.
In the 3D geometry in Fig.~\ref{Fig: TI surface}(b),
massive Dirac fermions
show quantum anomalous Hall effect
$\sigma_{xy}=-\t{sgn}(Jm)e^2/2h$ and $\sigma_{xy}=\t{sgn}(Jm)e^2/2h$
at the top and bottom surfaces,
respectively.
Since the electron density on each surface is given by
$n=\sigma_{xy}B$ ,
applying a magnetic field induces the charge polarization
\begin{align}
P&=
\frac{1}{2d} \left\{
d\!\left[-\textrm{sgn}(Jm)\frac{e^2}{2h} \right] B
+(-d)\textrm{sgn}(Jm)\frac{e^2}{2h} B
\right\}
\n
&=-\t{sgn}(Jm) \frac{e^2}{2h} B.
\label{eq: P in magnetic TIs}
\end{align}%
Thus the TME effect with the axion angle $\theta=-\pi\,\t{sgn}(Jm)$
is realized with a uniform magnetization.
Since Dirac fermions on the side surface are gapped
because of the finite thickness, the TME effect should be observed
when the chemical potential is tuned into the energy gap
of the side-surface excitations.
Compared with the original proposal with the non-uniform
configuration of magnetization \cite{Nomura}
which is experimentally difficult to realize,
this setup has the advantage that it only requires a uniform magnetization,
which facilitates experimental realization.

For both setups
the charge polarization can be detected as a charge current
in response to time-dependent $B$.

\section{Non-topological contributions from the bulk and the side surface}
In general, the axion angle can deviate from 
$\theta=0,\pi$ (mod $2\pi$)
when both time reversal symmetry $T$ and inversion symmetry $P$ are broken.
The $\nu=0$ QHE is achieved in TIs by applying magnetic field,
which breaks $T$, and by inducing potential asymmetry
between the top and bottom surfaces, which breaks $P$.
Thus the axion angle in the $\nu=0$ QHE setup is not necessarily quantized.
In the previous section we have shown that polarization corresponding
to the quantized value of $\theta=\pi$ (mod $2\pi$) is obtained
when we neglect contributions from the bulk 
and focus on contributions from the top and bottom surfaces.
While the charge polarization in the TME effect 
is accompanied by a flow of the surface current,
non-topological contributions may also arise from the current flowing
through the bulk.
Furthermore, the side surfaces connecting the top and bottom
surfaces in Fig.~\ref{Fig: TI surface}
may give rise to non-topological contributions. 
Then the natural question is whether
charge polarization of TI thin films is really quantized as
Eq.\ (\ref{P = theta B}) predicts with the axion angle $\theta=\pi$ (mod $2\pi$)
when contributions from the bulk and the side surface are taken into account.
To address this question,
we consider the 3D bulk of TI thin films and
calculate the charge polarization and the current distribution
induced by increasing $B$.

\begin{figure}[tb]
\begin{center}
\includegraphics[width=0.8\linewidth]{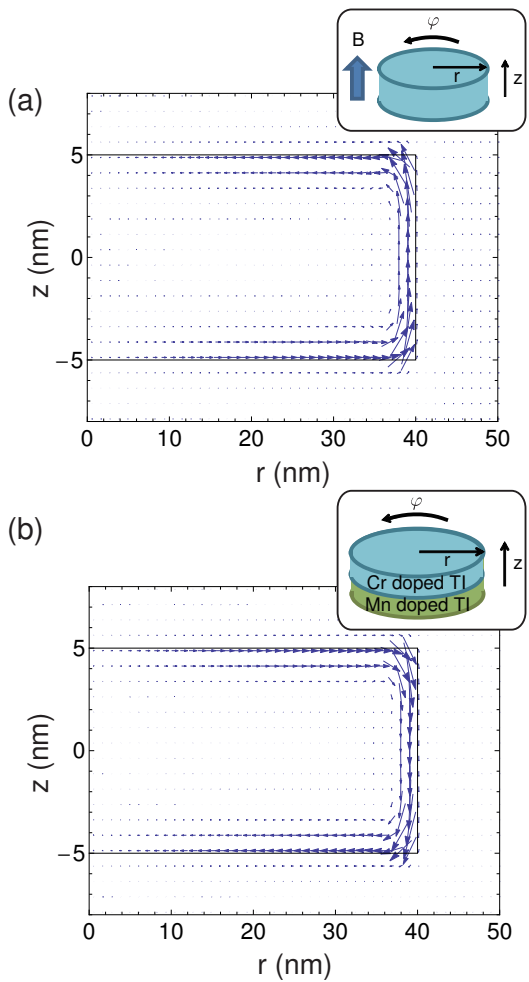}
\end{center}
\caption{
Current density (indicated by arrows) induced by small time-dependent
magnetic field $J(\Vec r)/(d\delta B/dt)$
for (a) TIs in the $\nu=0$ QHE with  $B= 10 \t{T}, U_0=20 \t{meV},$
and (b) magnetically doped TIs with $M_0=10 \t{meV}$. 
Insets are schematic pictures of the cylinder geometry of TI films.
We adopted parameters:
$v_F=5\times 10^5~\t{m/s}$, $2d=10~\t{nm}$, $R_0=40~\t{nm}$,
$m_\t{in}=250~\t{meV}$, and $m_\t{out}=-750~\t{meV}$.
}
\label{Fig: current density}
\end{figure}

Suppose that the TI sample has the cylinder geometry shown
in Fig.~\ref{Fig: current density},
in which the top and bottom surfaces are located at $z=\pm d$ and 
the side surface is at $r=R_0$.
In order to describe the bulk electronic states of the TI film,
we take a simple massive Dirac Hamiltonian
\begin{align}
H&=H_0+V, &
H_0&=v_F(\Vec{p} \cdot \Vec{\sigma})\tau_x + m(\Vec{r})\tau_z, 
\label{eq: H 3D bulk}
\end{align}
where
$\Vec{p}=(p_x,p_y,p_z)$ is the momentum, and
$\Vec{\sigma}=(\sigma_x,\sigma_y,\sigma_z)$
is the spin Pauli matrices, 
while $\tau_x, \tau_z$ are the orbital Pauli matrices.
We assume that the chemical potential $\mu=0$.
The Dirac mass (i.e., the bulk band gap)
$m(\Vec{r})=m_\t{in}>0$ inside the TI film.
The Dirac mass $m(\bm{r})$ changes its sign at the surface,
and $m(\Vec{r})=m_\t{out}<0$ outside the TI.
In the setup of the $\nu=0$ QHE in Fig.~1(a), $V$ is given by
\begin{align}
V(\Vec{r})= v_F e \Vec{A} \cdot \Vec{\sigma} \tau_x +U(z), 
\end{align}
where $\Vec A$ is the vector potential in the symmetric gauge 
$\Vec{A}=\frac{1}{2}B(-y,x,0)$,
and the potential energy $U(z)=U_0$
for $z\ge0$ and $U(z)=-U_0$ for $z<0$.
In the setup of a TI doped with Cr and Mn in Fig.~1(b), $V$ is given by 
\begin{align}
V(\Vec{r})=M(z)\sigma_z,
\end{align}
where $M(z)=M_0$ for $z\ge0$ and $M(z)=-M_0$ for $z<0$.
In the cylindrical coordinates $(r,\varphi,z)$,
wave functions are written as
\begin{align}
\psi(r,\varphi,z)&=
r^{-\frac{1}{2}} e^{i \left(\ell-\frac{\sigma_z}{2}\right) \varphi}
\phi_\ell(r,z)
\end{align}
for the angular momentum $\ell \in \mathbb{Z}+\frac{1}{2}$.
Then the Dirac equation is reduced to 
$[H_0(\ell)+V] \phi_\ell = E(\ell) \phi_\ell$, where
\begin{align}
H_0(\ell) =& 
v_F \!\left(
p_r \sigma_x + \frac{\hbar \ell}{r} \sigma_y + p_z \sigma_z 
\right)\! \tau_x +  m(\Vec{r})\tau_z,
\end{align}
and
$\phi(r,z)$ obeys the Dirichlet boundary condition $\phi=0$ at $r=0$.

Small magnetic field $\delta B(t)$ applied in the $z$ direction
(in addition to the constant magnetic field $B$ for the $\nu=0$ QHE case)
induces a change in the polarization along the $z$ direction
that is proportional to $\delta B(t)$.
The change in the polarization leads to
polarization current $\bm{J}$ proportional to 
$d\delta B(t)/dt$.
From the linear response theory
(see Appendix~\ref{app: linear response theory} for details),
the induced current density is written as
\begin{align}
\frac{\Vec{J}(\Vec{r}_0)}{d\delta B/dt}=&
\sum_{
\scriptsize
\begin{array}{c}
i \in O \\ 
j \in U 
\end{array}
}
\frac{2\hbar~
\t{Im} \left[
\bra{\psi_{i}} \Vec{j}(\Vec{r}_0) \ket{\psi_{j}}
\bra{\psi_{j}} {\cal O} \ket{\psi_{i}}\right]
}{(E_i-E_j)^2} 
,
\label{eq: current density}
\end{align}
where $i$ and $j$ label occupied states $\psi_i$ ($i\in O$)
and unoccupied states $\psi_j$ ($j\in U$)
with energies $E_i$ and $E_j$, respectively, 
$\Vec{j}(\Vec{r_0})= - e v_F \delta(\Vec{\hat r}-\Vec{r}_0)
 \Vec{\sigma}\tau_x$,
and ${\cal O}=\frac{1}{2}e v_F (-y\sigma_x + x\sigma_y) \tau_x$.

The current density in Fig.~\ref{Fig: current density} shows that
the current flows along the surface of the TI film,
not through the bulk.
Thus the charge polarization in the present setup is indeed
a topological phenomenon carried by the Dirac fermions
on the surface of TIs.
The surface current of topological nature is
understood as follows.
Time-dependent $\delta B$ induces an electric field
along the azimuthal ($\varphi$) direction.
Then the Hall current flows along the $r$-direction
on the top and bottom surfaces.
The directions of the Hall current are opposite for the top and bottom
surfaces, where $\sigma_{xy}=\pm e^2/2h$ and $\sigma_{xy}=\mp e^2/2h$,
respectively.
On the side surface, the current in the $z$ direction flows
to connect current distributions at the top and bottom surfaces,
and it is exactly the polarization current along the $z$ direction 
in the TME effect.
From the discussions focusing on the top and bottom surfaces
in the previous section,
the polarization current is expected to be equal to the time derivative
of $\bm{P}$ in Eq.\ (\ref{P = theta B}) with the axion angle at
$\theta=\pi$ for the $\nu=0$ QHE from Eq.~(\ref{eq: P in nu=0 QHE}) with $N=0$,
and at $\theta=-\pi$ for the magnetically doped TI
from Eq.~(\ref{eq: P in magnetic TIs}) with $Jm=M_0>0$.
However, the axion angles obtained from the integration of
the polarization current plotted in Fig.~2,
\begin{equation}
\frac{\theta}{2\pi}
=\frac{h}{e^2} \frac{1}{\pi R_0^2 (d\delta B/dt)}
\int r dr d\theta J_z(z=0),
\end{equation}
are found to be $0.4$ and $-0.3$ for the $\nu=0$ QHE
and the magnetically doped TIs, respectively,
and both deviate from the quantized value $\theta/2\pi= \pm\frac12$.
We attribute these deviations to finite-size effects
of the surface of TI films,
considering that the bulk contribution to the polarization current is
negligibly small as shown in Fig.~\ref{Fig: current density}(a) and (b).
Specifically, Dirac fermions on the side surface
give non-topological contributions
to the polarization current
because of broken time-reversal and inversion symmetries.

\section{Analysis from surface Weyl fermions and the quantized axion angle}

\begin{figure}[tb]
\begin{center}
\includegraphics[width=0.8\linewidth]{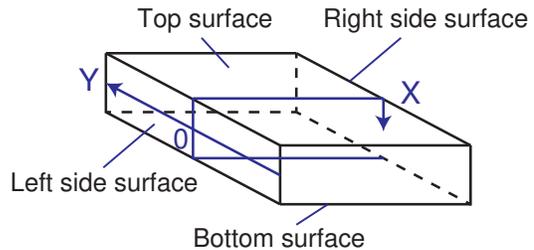}
\end{center}
\caption{
Schematic picture of the effective 2D model with coordinates
$(X,Y)$ along the surface of a TI thin film.
}
\label{Fig: 2D coordinate}
\end{figure}

The current distribution in Fig.~\ref{Fig: current density}
indicates that the polarization current is negligible in the gapped bulk
and flows only at the surface of TI thin films.
We now focus on low-energy Dirac fermions at the surface 
and study their contribution to the magneto-electric effect.
In particular, we are interested in finite-size corrections
which cause the deviation of $\theta$
from the quantized value.
For this purpose, we introduce an effective 2D model
(Fig.~\ref{Fig: 2D coordinate}) for
surface Dirac fermions,
in which surfaces of a 3D TI thin film are extended
on the 2D $(X,Y)$ plane.
The whole 2D surface has linear dimensions of $(2L_x+4d) L_y$ and 
the $y$ direction is assumed to be periodic.
The original 3D coordinates $(x, y, z)$ of each surface
are related to 2D coordinates $(X,Y)$ as follows:
\begin{align}
(x, y, z)&\!=\!
\begin{cases}
(-X + d + \frac{L_x}{2},Y, -d), 
&-L_x-d<X<-d, \\
(X - d - \frac{L_x}{2},Y, d), 
&d<X<L_x+d, \\
(-\frac{L_x}{2},Y,X), &|X| < d, \\
(\frac{L_x}{2}, Y, L_x+d-X), & L_x+d < X, \\
(\frac{L_x}{2}, Y, -L_x-d-X), & X<-L_x-d, 
\end{cases}
\end{align}
see Fig.~\ref{Fig: 2D coordinate}.

\begin{figure}[tb]
\begin{center}
\includegraphics[width=0.7\linewidth]{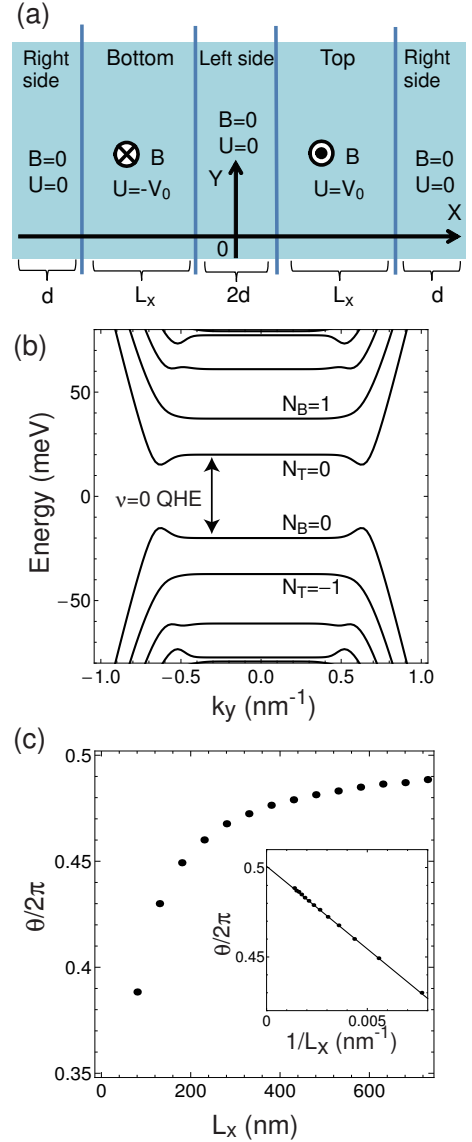}
\end{center}
\caption{
(a) Schematic picture of a TI thin film under magnetic fields.
(b) Band structure of a TI thin film for $L_x=200 \t{nm}$. 
(c) Axion angle in the $\nu=0$ QHE plotted against the system size $L_x$.
The inset shows axion angle plotted against the inverse of the system size
$1/L_x$.
We performed the calculation with parameters:
$v_F=5\times 10^5 \t{m/s}$, $B=10 \t{T}$, $U_0=20 \t{meV}$, and
$2d=10 \t{nm}$.
}
\label{Fig: QHS}
\end{figure}

For the $\nu=0$ QHE setup
the effective Hamiltonian for the surface Dirac fermions
is given by
\begin{align}
H= v_F\{ -p_X \sigma_y + [p_Y+ e A_Y(X)] \sigma_x\} + U(X) \sigma_0.
\label{eq: Dirac Hamiltonian}
\end{align}
Here, 
we take the Landau gauge $A_Y(X)= B x(X)$.
The potential difference between the top and
bottom surfaces is given by the potential
\begin{equation}
U(X)=
\begin{cases}
U_0, & d<X<L_x+d,\\
-U_0, & -L_x-d<X<-d,\\
0, & \mathrm{otherwise},
\end{cases}
\end{equation} %
as schematically illustrated in Fig.~\ref{Fig: QHS}(a).
Since the 2D model is translationally invariant along the $Y$ direction
under the Landau gauge,
the wave number $k_Y (=k_y)$ is a good quantum number.

The polarization in the $z$ direction is given by
\begin{align}
P=\frac{-e}{2dL_x}z(X)
.
\end{align} %
We note that the spectrum of $z$ is bounded in our 2D effective model
so that the above equation is well defined.
From the linear-response theory,
the axion angle $\theta$ determined from the charge polarization 
$\theta/2\pi=h\delta P/e^2 \delta B$
is given by 
\begin{align}
\frac{\theta}{2\pi}
&=
\frac{-e}{2d L_x} 
\frac{h}{e^2}
\int \frac{d k_y}{2\pi}
\sum_{
\scriptsize
\begin{array}{c}
i \in O \\ 
j \in U 
\end{array}
}
\frac{2}{E_{i,k_y}-E_{j,k_y}}
\n
&~\quad\times
\t{Re}[
\bra{\psi_{i,k_y}} z \ket{\psi_{j,k_y}} 
\bra{\psi_{j,k_y}} e v_F x \sigma_x \ket{\psi_{i,k_y}}
]
,
\label{eq: linear response}
\end{align}
where $i$ and $j$ label occupied states $\psi_{i,k_y}$ ($i\in O$)
and unoccupied states $\psi_{j,k_y}$ ($j\in U$)
with momentum $k_y$ and energy $E_{i,k_y}$ and $E_{j,k_y}$, respectively.
To compute Eq.\ (\ref{eq: linear response}) numerically,
we discretize the 2D model.
In order to avoid the fermion doubling
on the lattice,
we used a regularization scheme in the momentum space. 
Namely, we used the plane wave basis for the $X$ direction labeled
by $k_X$ and set a cutoff to the wave number $k_X$ as $|k_X|< \Lambda$.
The numerical results are obtained by extrapolating the cutoff
$\Lambda \to \infinity$.

We show results for the axion angle computed from
Eq.\ (\ref{eq: linear response}) in Fig.~\ref{Fig: QHS}.
The axion angle $\theta/2\pi$ approaches $1/2$ as $L_x\to \infinity$.
This shows that non-topological contributions from the side surface
become negligible as the ratio $L_x/d$ is increased.
Indeed the deviation of the computed axion angle $\theta/2\pi$
from the quantized value $1/2$
scales with the ratio of the areas of the side surface and
the top/bottom surfaces, $d/L_x$, 
as shown in Fig.~\ref{Fig: QHS}(c).
Thus the axion angle converges to the quantized value $1/2$ 
in the limit where the top/bottom surfaces have much larger
area than side surfaces.
Since typical values of the thickness $2d$ and the size $L_x$ of
TI films are
$2d=10\t{ nm}$ and $L_x=1\t{ mm}$ \cite{Yoshimi},
the axion angle $\theta/2\pi$ is expected to show a quantized behavior
in practice.

\begin{figure}[tb]
\begin{center}
\includegraphics[width=0.7\linewidth]{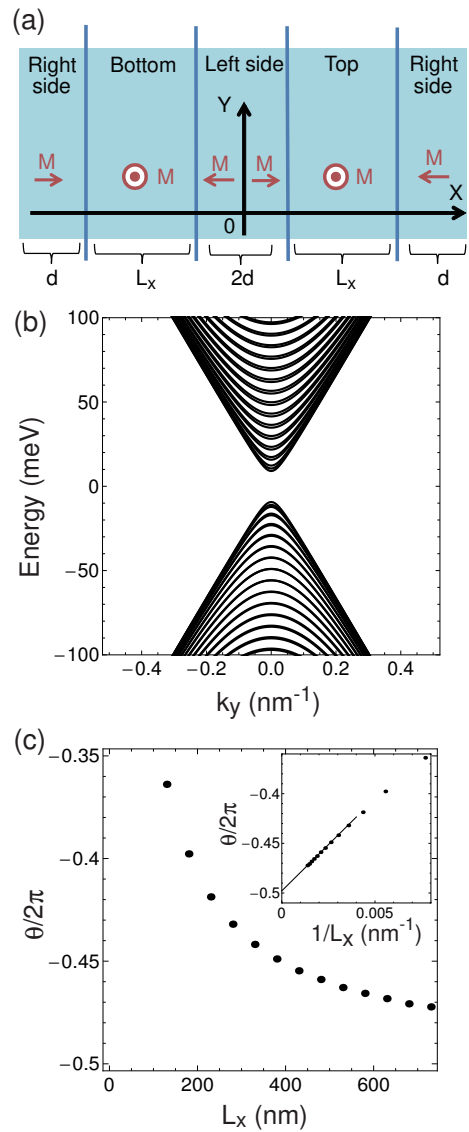}
\end{center}
\caption{
(a) Schematic picture of a TI thin film doped with magnetic ions, Cr and Mn.
(b) Band structure of a TI thin film for $L_x=200 \t{nm}$. 
(c) Axion angle plotted against the system size $L_x$.
The inset shows axion angle plotted against the inverse of the system size
$1/L_x$.
We performed the calculation with parameters:
$v_F=5\times 10^5 \t{m/s}$, $M_0=10 \t{meV}$, and $2d=10 \t{nm}$.
}
\label{Fig: ferromagnets}
\end{figure}

Next we move on to the 2D surface model for the magnetically
doped TI.
The effective Hamiltonian for the surface Weyl fermions is given by
\begin{align}
H= v_F( -p_X \sigma_y + p_Y \sigma_x)
 + \Vec{M} \cdot \Vec{\sigma},
\label{eq: Dirac Hamiltonian 2}
\end{align}
where we assumed no potential asymmetry between the top and bottom surfaces.
The upper and lower halves of the TI film are doped with Cr and Mn,
respectively.
A uniform magnetization along the $z$ direction induces effective
exchange fields $+M_0$ and $-M_0$ along the $z$ direction
in the upper and lower halves of the TI film, respectively,
because signs of exchange couplings are opposite for Cr and Mn. 
In the 2D model of surface Weyl fermions, these exchange fields are
perpendicular to the top and bottom surfaces
\begin{equation}
\Vec M(X)=(0, 0, M_0)
\end{equation}
for $|X \pm (L_x/2 +d)| < L_x/2$
and produce a mass term $M_0\sigma_z$, 
while they are perpendicular to the side surface
\begin{equation}
\Vec M(X)=(\pm M_0, 0, 0)
\end{equation}
for $-d<\pm X+L_x+d<0$ and $0<\pm X< d$
and enter as a vector potential $\pm M_0\sigma_x$
in Eq.\ (\ref{eq: Dirac Hamiltonian 2}).
This is schematically illustrated in Fig.~\ref{Fig: ferromagnets}(a).
We numerically obtain the polarization and the axion angle
using the linear response theory [Eq.~(\ref{eq: linear response})]
in the same manner as in the $\nu=0$ QHE case.
The result is shown in Fig.~\ref{Fig: ferromagnets}.
We again see asymptotic behavior $\theta/2\pi \to -1/2$
with $L_x\to \infinity$ for fixed $d$.
In a similar way to the case of the $\nu=0$ QHE, 
the deviation of the axion angle from the quantized value  $-1/2$
scales with $d/L_x$ as shown in Fig.~\ref{Fig: ferromagnets}(d).
Thus there is a non-topological contribution 
proportional to the area of the side surface,
but it diminishes in the thermodynamic limit $L_x/d\to\infty$
where typical experiments of thin films are performed.

\section{Discussions}

\begin{figure}[tb]
\begin{center}
\includegraphics[width=0.6\linewidth]{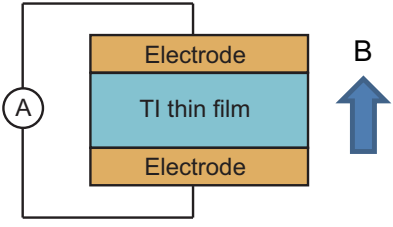}
\end{center}
\caption{
A setup to observe the TME effect in TI thin films by detecting
the polarization current induced by changing magnetic fields. 
}
\label{Fig: detection setup}
\end{figure}

We have proposed that the TME effect should be observed in TI films
in the $\nu=0$ QHE and in magnetically doped TI films.
We have shown that the polarization current in these setups are topological
in that bulk contributions are negligible and show a quantized response in
the limit of large top/bottom surface area.
The non-topological contribution to the polarization current is proportional
to the area of the side surface,
which is interpreted as the axion angle $\theta$ at the side surface
deviating from the quantized value $\pm \pi$ due to the breaking of
time-reversal and inversion symmetries.
For a thin-film sample of area $1\t{mm}^2$ and 
a $B$-field sweeping speed of $1\t{T}/\t{s}$,
the polarization current is estimated to be $20 \t{pA}$.
This magnitude of the polarization current is accessible in
the current state of transport measurements \cite{Omori14}.
Application of ac magnetic fields induces a polarization current
oscillating in time in the $z$-direction.
This ac polarization current can be detected, for example,
with electrodes capacitively coupled
to top and bottom surfaces
and measuring currents between two electrodes
as schematically shown in Fig.~\ref{Fig: detection setup}.
Besides the corrections to the TME effect [Eq.~(\ref{P = theta B})]
due to broken $T$ and $P$ symmetries in our setup,
there can be other corrections from
other mechanisms, such as the coupling between top and bottom surfaces.
However, the dominant part of the polarization should come from
the topological contribution.

In the ideal situation where the axion action in Eq.~(\ref{eq:theta})
with $\theta=\pi$ is realized over the whole sample, 
the total charge induced by $\Vec{B}$ is proportional to the cross
sectional area of the sample perpendicular to $\Vec{B}$ as follows.
We suppose that $\Vec{B}$ is parallel to the $z$ direction,
and the sample geometry
is defined by height functions for upper and lower surfaces, 
$z=z_u(x,y)$ and $z=z_l(x,y)$, respectively,
for $x,y$ within the projected region $S$ of the sample onto the $xy$-plane.
The total charge on the upper surface is given by
\begin{align}
Q &=
\int_S dx dy
\frac{1}{z_u(x,y)-z_l(x,y)}
\int^{z_u(x,y)}_{z_l(x,y)} dz \frac{\theta}{2\pi}\frac{e^2}{h} B 
\n
&=
\frac{\theta}{2\pi}\frac{e^2}{h} B A_S,
\end{align}
where $A_S=\int_S dx dy$ is the cross sectional area of the sample
in the $xy$-plane.
Conversely,
the axion angle $\theta$ deviates from the quantized value 
in our setups for the TME effect,
since there is
non-topological contribution proportional to the area of the side surface
 as shown in Fig.~\ref{Fig: QHS}(c) and Fig.~\ref{Fig: ferromagnets}(c).

\textit{Note added:} After the completion of this work, a paper on a related topic by Wang et al. \cite{wang2015quantized} has appeared.

\begin{acknowledgments}
We thank Y. Tokura and M. Kawasaki for fruitful discussions.
This work was supported by 
Grant-in-Aid for Scientific Research 
(No.~24224009, No.~26103006, No.~24540338, and No.~15K05141)
from the Ministry 
of Education, Culture, Sports, Science and
Technology (MEXT) of Japan
and from Japan Society for the Promotion of Science.
\end{acknowledgments}

\appendix*

\section{Linear response theory for current density and polarization
\label{app: linear response theory}}

We give a brief derivation of Eqs.\ (\ref{eq: current density})
and (\ref{eq: linear response})
from linear response theory.
The Hamiltonian
\begin{align}
H=H_0 + \mathcal{A}(t),
\end{align}
consists of an unperturbed part $H_0$
and a perturbation
$\mathcal{A}(t)\equiv \mathcal{A} e^{-i\omega t}$.
We are interested in the expectation value of an operator
$\mathcal{B}$ in the ground state of $H$,
\begin{align}
\bracket{\mathcal{B}}=\mathcal{B}_\omega e^{-i\omega t},
\end{align}
which is expanded in powers of $\mathcal{A}$.
The Fourier component $\mathcal{B}_\omega$ in linear order in $\mathcal{A}$
is obtained from the linear response theory,
\begin{align}
\mathcal{B}_\omega=
-\frac{i}{\hbar} \int_0^\infinity d\tau e^{i\omega \tau} \t{tr}
\!\left[
e^{-iH_0\tau/\hbar} [\mathcal{A},\rho_0] e^{iH_0\tau/\hbar} \mathcal{B}
\right] ,
\label{B_omega linear in A}
\end{align}
where the density matrix $\rho_0=|0\rangle\langle0|$
is the projection to the many-body ground state $|0\rangle$ of $H_0$.
For noninteracting fermions, $\rho_0$ can be written as
$\rho_0=\prod_{i\in O} \ket{\psi_i} \bra{\psi_i}$
in terms of the occupied single-particle states $\psi_i$.
The Fourier component $\mathcal{B}_\omega$
in Eq.\ (\ref{B_omega linear in A}) is then rewritten as
\begin{align}
\mathcal{B}_\omega &=
-\frac{i}{\hbar} \int_0^\infinity d\tau \!
\sum_{
\scriptsize
\begin{array}{c}
i \in O \\ 
j \in U 
\end{array}
}
\!\left[
\mathcal{B}_{ij} \mathcal{A}_{ji}
e^{i(E_i-E_j+\hbar \omega) \tau/\hbar} \right. \n
& \hspace{9em} \left.
-\mathcal{B}_{ji} \mathcal{A}_{ij}
e^{i(E_j-E_i+\hbar \omega) \tau/\hbar} \right],
\n
&=
\sum_{
\scriptsize
\begin{array}{c}
i \in O \\ 
j \in U 
\end{array}
}\!\!
\left[
\frac{\mathcal{B}_{ij} \mathcal{A}_{ji}}{E_i-E_j+\hbar \omega} + 
\frac{(\mathcal{B}_{ij} \mathcal{A}_{ji})^*}{E_i-E_j-\hbar \omega}
\right]
\n
&=
\sum_{
\scriptsize
\begin{array}{c}
i \in O \\ 
j \in U 
\end{array}
}\!\!
\frac{2
[
(E_i-E_j)\t{Re}(\mathcal{B}_{ij} \mathcal{A}_{ji}) 
-i \hbar \omega \t{Im}(\mathcal{B}_{ij} \mathcal{A}_{ji})
]
}{(E_i-E_j)^2-(\hbar\omega)^2}
,
\label{B_omega}
\end{align}
where $O$ and $U$ stand for sets of occupied and unoccupied single-particle
states in the ground state of $H_0$, and
$\mathcal{A}_{ij}$ and $\mathcal{B}_{ij}$ are matrix elements of
the operators $\mathcal{A}$ and $\mathcal{B}$.
We have used the hermiticity of the operators $\mathcal{A}$ and $\mathcal{B}$
when passing from the first line to the second line.
In the dc limit Eq.\ (\ref{B_omega}) reduces to
\begin{align}
\mathcal{B}_{\omega \to 0} =
\sum_{
\scriptsize
\begin{array}{c}
i \in O \\ 
j \in U 
\end{array}
}
\frac{2 \, \t{Re}(\mathcal{B}_{ij} \mathcal{A}_{ji}) }{E_i-E_j}.
\label{B_omega_to_0}
\end{align}
The axion angle $\theta$ is defined in Sec.~IV as a response coefficient
of polarization $P$ to magnetic field $B$.
The linear response formula [Eq.~(\ref{eq: linear response})]
is obtained by setting $\mathcal{B}=P$ and
$\mathcal{A}=ev_F x \sigma_x B$ in Eq.\ (\ref{B_omega_to_0}).

In order to obtain the formula in Eq.~(\ref{eq: current density})
for the current density that is proportional to the time derivative
of the magnetic field,
we need to extract from $\langle\mathcal{B}\rangle$ the component
that is proportional to the time derivative of the perturbation,
$d\mathcal{A}(t)/dt=-i\omega \mathcal{A} e^{-i\omega t}$.
That component is given by
\begin{align}
\lim_{\omega\to 0} \t{Re}[\mathcal{B}_\omega/(-i\omega)]
&=
\sum_{
\scriptsize
\begin{array}{c}
i \in O \\ 
j \in U 
\end{array}
}
\frac{2 \hbar\, \t{Im}(\mathcal{B}_{ij} \mathcal{A}_{ji}) }{(E_i-E_j)^2}.
\label{B/(-i omega)}
\end{align}
When the magnetic field is varied as $B \to B +\delta B$,
the 3D bulk Hamiltonian $H$ in Eq.~(\ref{eq: H 3D bulk}) is perturbed
by $\delta B \,{\cal O}=\frac{1}{2}ev_F(-y\sigma_x+x\sigma_y)\tau_x \delta B$.
Equation (\ref{eq: current density}) is obtained by setting
$\mathcal{B}=\Vec{j}$ and $\mathcal{A}={\cal O}$ in Eq.\ (\ref{B/(-i omega)}).

\bibliography{TME}

\end{document}